\documentclass[11pt, a4paper]{article}

\usepackage[utf8]{inputenc} 
\usepackage[T1]{fontenc}    
\usepackage{geometry}       
\usepackage{csquotes}
\usepackage{amsmath, amssymb}

\usepackage{graphicx}       
\usepackage{booktabs}       
\usepackage{longtable}      
\usepackage{xcolor}         

\usepackage{tcolorbox}
\tcbuselibrary{listings, skins, breakable}

\usepackage[colorlinks=true, allcolors=blue]{hyperref}

\usepackage{lipsum} 

\usepackage[english]{babel}
\usepackage{longtable}
\usepackage{booktabs} 
\usepackage{caption}  
\usepackage[
  backend=biber,
  style=numeric,
  sorting=none
]{biblatex}
\usepackage{listings}
\usepackage{xcolor}

\usepackage{authblk}
\usepackage{microtype}
\title{Brazilian Social Media Anti-vaccine Information Disorder Dataset - Telegram (2020-2025)}
\author[1]{João Phillipe Cardenuto\thanks{Corresponding Author}}
\author[1]{Ana Carolina Monari}
\author[1,2]{Michelle Diniz Lopes}
\author[1,3]{Leopoldo Lusquino Filho}
\author[1]{Anderson Rocha}

\affil[1]{Artificial Intelligence Lab., Recod.ai, Universidade Estadual de Campinas (UNICAMP), Brazil}
\affil[2]{Neuroscience Department, Federal University of Minas Gerais, Belo Horizonte, Brazil}
\affil[3]{São Paulo State University, Sorocaba, Brazil}

\date{}

\begin{document}

\maketitle

\begin{abstract}
\noindent

Over the past decade, Brazil has experienced a decline in vaccination coverage, reversing decades of public health progress achieved through the National Immunization Program (PNI). Growing evidence points to the widespread circulation of vaccine-related misinformation—particularly on social media platforms—as a key factor driving this decline. Among these platforms, Telegram remains the only major platform permitting accessible and ethical data collection, offering insight into public channels where vaccine misinformation circulates extensively.
This data paper introduces a curated dataset of about four million Telegram posts collected from 119 prominent Brazilian anti-vaccine channels between 2020 and 2025. The dataset includes message content, metadata, associated media, and classification related to vaccine posts, enabling researchers to examine how false or misleading information spreads, evolves, and influences public sentiment. By providing this resource, our aim is to support the scientific and public health community in developing evidence-based strategies to counter misinformation, promote trust in vaccination, and engage compassionately with individuals and communities affected by false narratives. The dataset and documentation are openly available for non-commercial research, under strict ethical and privacy guidelines at \href{https://doi.org/10.25824/redu/5JIVDT}{https://doi.org/10.25824/redu/5JIVDT}.
\\
\textbf{Keywords:} Vaccine Misinformation, Information Disorder, Social Media, Telegram, Public Health, Dataset.
\end{abstract}

\section{Introduction}
\label{sec:intro}

For nearly half a century, Brazil's National Immunization Program (Programa Nacional de Imunizações - PNI) has stood as a standard of public health achievement, not only within Latin America but across the globe~\cite{Massuda2018,ministerio2003pni}. Established in 1973, the PNI was conceived as a cornerstone of the nation's public health strategy, designed to coordinate and execute vaccination actions with the goal of preventing disease and elevating the quality of life for the entire population. Operating through the federally funded public health system, the \textit{Sistema Único de Saúde} (SUS), the program has been instrumental in providing universal and free access to a complete calendar of vaccines, a reality not shared by many other nations~\cite{Possas2024}. It is estimated that the PNI is responsible for the administration of between 90\% and 95\% of all vaccines in Brazil, a demonstration of its central role in the nation's health infrastructure~\cite{Moura2022}.

The achievements of the PNI have been dangerously undermined by a progressive decline in vaccination coverage. A survey form the Health Ministry indicates that this downward trend took hold around 2020~\cite{ButantanHesitacaoVacinal}, marking a stark reversal of decades of public health progress.  The mean child vaccination coverage across the country, which reached respectable 97\% in 2015, dropped to just 75\% by 2020~\cite{ButantanHesitacaoVacinal}. 

The decline in Brazil's vaccination rates cannot be explained by logistical failures alone. A primary driver of this crisis is the complex and pervasive phenomenon of ``vaccine hesitancy,'' a term defined by the World Health Organization (WHO) as a delay in the acceptance or outright refusal of vaccines despite their availability~\cite{ButantanHesitacaoVacinal}. This hesitancy, which ranges from cautious deliberation to persistent opposition, has been identified by global health authorities as one of the most significant threats to public health and is a key factor contributing to falling immunization coverage in Brazil and worldwide~\cite{Borges2024}.

This growing hesitancy is fueled by a concurrent public health crisis: the ``infodemic.'' An infodemic is characterized by an overwhelming excess of information, much of it inaccurate, misleading, or maliciously false, that spreads with viral speed through digital channels, making it exceedingly difficult for the public to identify trustworthy sources~\cite{Cunha2025}. Numerous studies have established a direct link between the proliferation of this fake news\footnote{In this work, we consider fake news as one of the forms of Information Disorder~\cite{clairewardle2017}, comprising three categories: disinformation (false information created to cause harm), misinformation (false information without harmful intent), and malinformation (true information used to cause harm).} and the increase in vaccine hesitancy in Brazil~\cite{Pereira2024, Galhardi2022, Medeiros2022}.

A comprehensive national survey conducted in 2024 by the National Council of the Public Ministry (CNMP) revealed that at least one in five Brazilians has either felt fear about getting vaccinated or has decided against vaccination for themselves or a child under their care specifically after being exposed to negative news or misinformation on digital platforms~\cite{Galhardi2022}. This demonstrates that the digital information environment is not a passive backdrop but an active agent shaping health-related decisions and behaviors at a national scale.

Motivated by this scenario, this data paper presents a curated dataset of Telegram posts collected from Brazilian anti-vaccine channels. The dataset is intended to support researchers, health communicators, and public institutions in understanding the patterns and narratives that sustain vaccine misinformation in digital spaces. By making these data openly available, we aim to facilitate multidisciplinary studies capable of revealing how misinformation spreads, evolves, and affects public perceptions. More importantly, this work seeks to contribute to the development of informed and empathetic public health strategies-ones that not only counter false narratives but also rebuild trust, compassion, and dialogue with individuals and communities who have been most exposed to and influenced by such misinformation.

The remainder of this paper is organized as follows: Section~\ref{sec:preliminary} presents a preliminary analysis of vaccine-related misinformation; Section~\ref{sec:collection} describes the data collection and processing methods; Section~\ref{sec:description} outlines the structure and schema of the dataset; Section~\ref{sec:uses} discusses potential applications and research uses of the data; Section~\ref{sec:availability} provides information about the public repository and ethical considerations for data access; Section~\ref{sec:limitation} examines the challenges and limitations encountered during data gathering; and finally, Section~\ref{sec:conclusion} concludes the paper.

\section{Preliminary Vaccine Fake News Analysis}
\label{sec:preliminary}
Prior to data collection, we analyzed the main themes of vaccine-related information disorder investigated by prominent Brazilian fact-checking agencies. For this purpose, we collected 164  articles published by these agencies in 2024 concerning fact-checked vaccine claims.

\begin{figure}[ht!]
    \centering
    \includegraphics[width=0.8\textwidth]{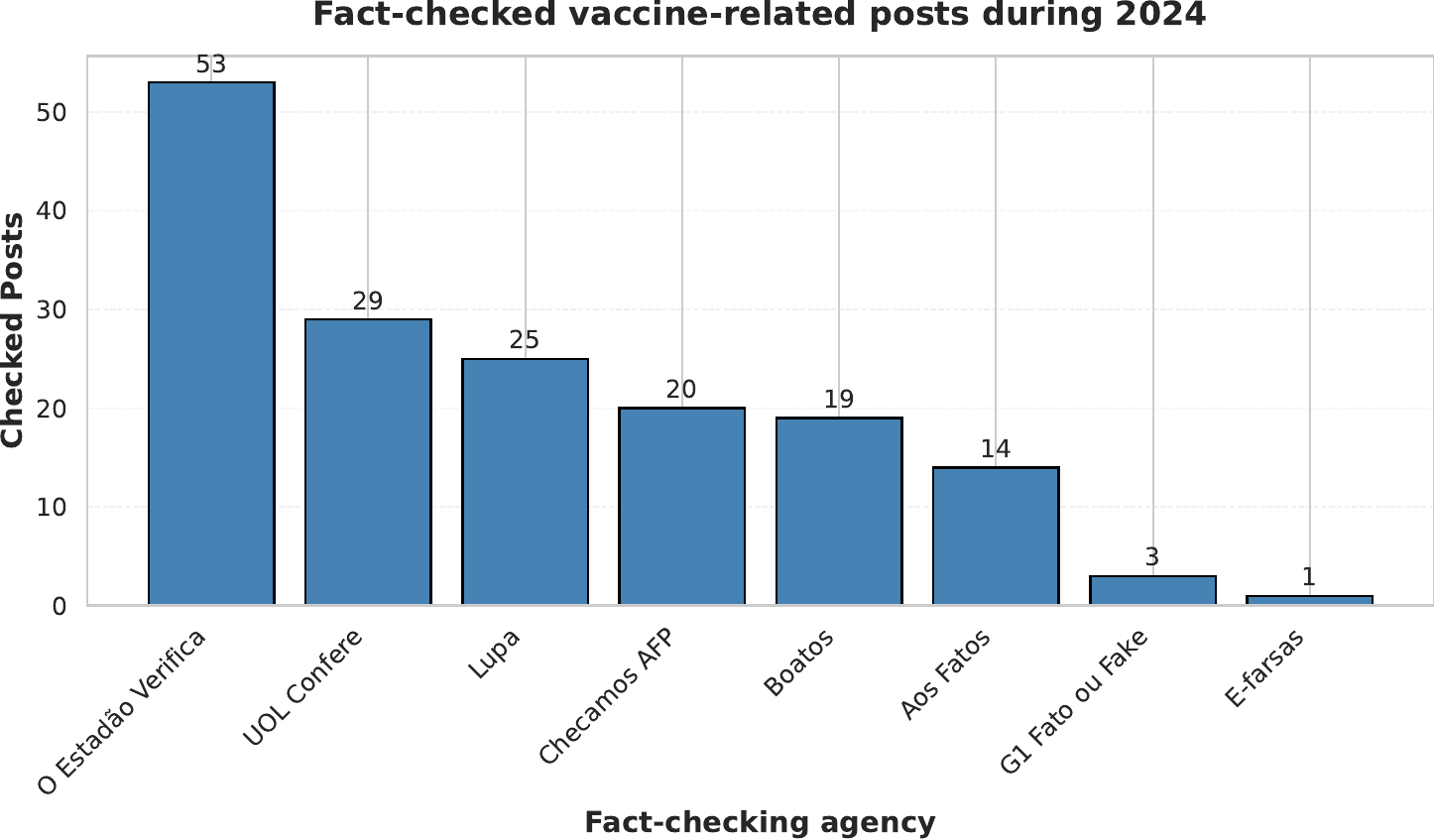}
    \caption{Distribution of fact-checking articles related to vaccines per agency published during 2024.}
    \label{fig:preliminary-fact-agencies}
\end{figure}

Figure~\ref{fig:preliminary-fact-agencies} indicates the number of articles each agency published during this period related to vaccine claims. \textit{Estadão Verifica} was the most proactive source, followed by \textit{UOL Confere} and \textit{LUPA}.

We utilized these articles to identify the primary fact-checked topics circulating in Brazil. The content of the articles was input into the Large Language Model (LLM) Sabiá-3~\cite{abonizio2024sabia3} from \hyperlink{https://www.maritaca.ai/}{Maritaca.AI}\footnote{\url{https://www.maritaca.ai/}}. This model is specialized in Brazilian culture and language, achieving high scores on Brazilian Portuguese benchmarks~\cite{abonizio2024sabia3}. Table~\ref{tab:preliminary:vac-topics} details the main topics identified by Sabiá-3 from the fact-checking agencies' articles.

\begin{longtable}{p{5cm} p{9.5cm}}
    \caption{2024 Fact-checked vaccine-related topics}
    \label{tab:preliminary:vac-topics} \\
    
    \toprule
    \textbf{Topic} & \textbf{Description} \\
    \midrule
    \endfirsthead
    
    \multicolumn{2}{c}%
    {{\tablename\ \thetable{} -- continued from previous page}} \\
    \toprule
    \textbf{Topic} & \textbf{Description} \\
    \midrule
    \endhead
    
    \bottomrule
    \endfoot
    
    \bottomrule
    \endlastfoot

    Claims of Severe Side Effects & False claims that COVID-19 vaccines cause side effects such as death, cancer, myocarditis, pericarditis, and neurodegenerative diseases. \\
    \addlinespace
    Distortion of Scientific/Medical Studies & Misleading information that questions the safety and efficacy of vaccines. Includes the distortion of scientific studies and the creation of false news about mRNA technology and other topics. \\
    \addlinespace
    Fraud and Falsification of Vaccination Cards & False reports alleging that authorities are involved in the circulation of fraudulent vaccination cards and the manipulation of vaccination data. \\
    \addlinespace
    Supposedly Genetic Effects & Unfounded claims that vaccines alter DNA or contain harmful genetic components. \\
    \addlinespace
    Disinformation on Government Policies & False stories related to vaccination policies, such as claims that the Brazilian government or other countries have issued false warnings or vaccine prohibitions. \\
    \addlinespace
    Conspiracy Theories & Theories suggesting that vaccines are intended to reduce the population, contain ``nanorobots,'' or are linked to political and global agendas. \\
    \addlinespace
    Distortion of Statements by Authorities and Specialists & Posts that take statements from doctors, health directors, and politicians out of context to create distrust in vaccines, or even manipulated content (deepfakes). \\
    \addlinespace
    Use of Alternative Medications & Dissemination of false information about drugs like ivermectin and suramin, suggesting they can reverse the effects of the vaccine or replace it in treating diseases. \\
    \addlinespace
    Comparative Disinformation & Erroneous comparisons about the safety and effects of vaccines with other diseases or events (e.g., Dengue or MPOX), creating confusion and fear. \\
    \addlinespace
    Political Campaigns and Disinformation & Candidates and politicians using platforms (e.g., Meta) to spread anti-vaccine disinformation. \\

\end{longtable}

The topics identified by Sabiá-3 indicate that the anti-vaccine discourse in Brazil encompasses a broader scope than vaccines alone, intersecting with other domains such as conspiracy theories, misinterpretation of scientific studies, and political polarization.

\begin{figure}[ht!]
    \centering
    \includegraphics[width=0.8\textwidth]{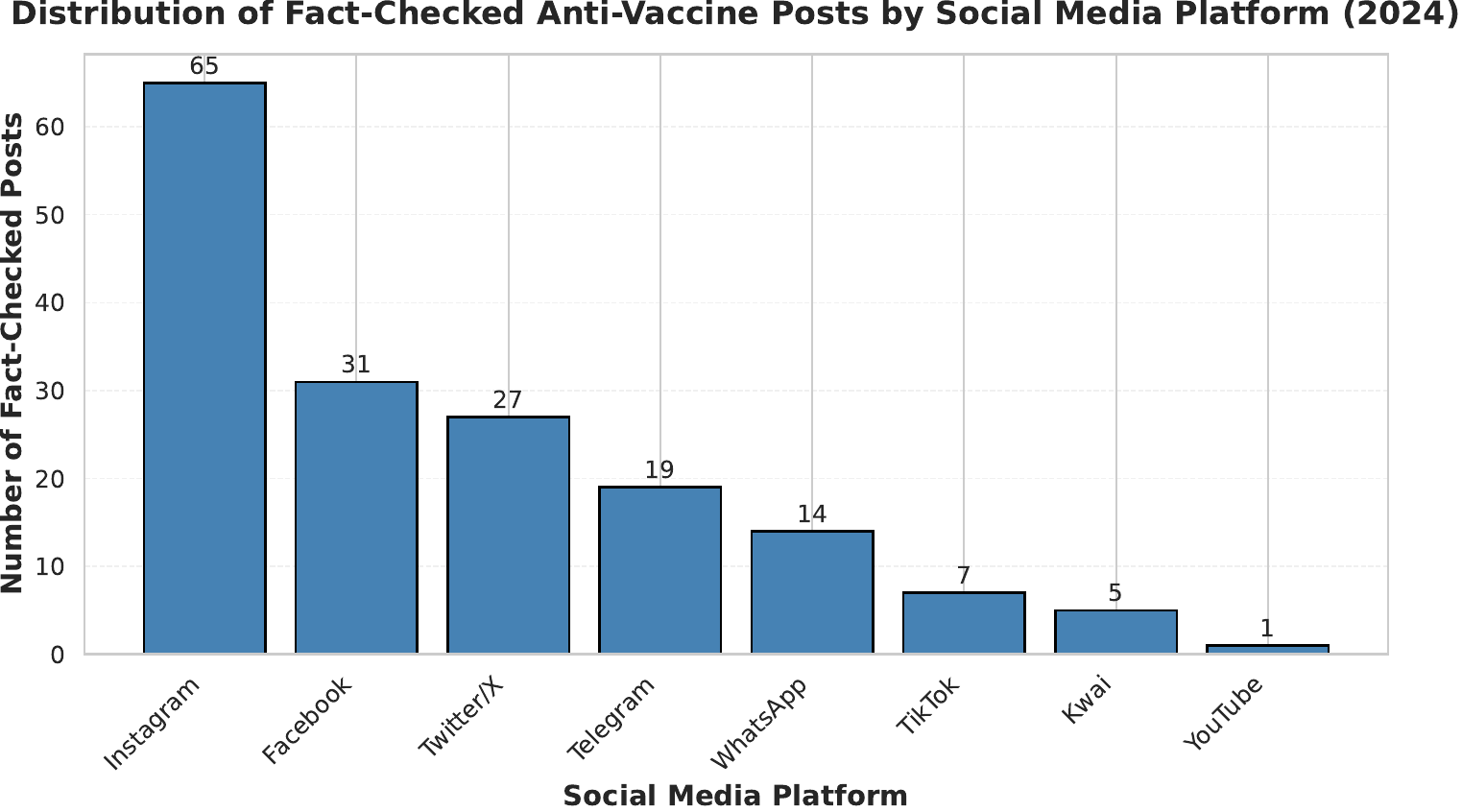}
    \caption{Distribution of reported Brazilian vaccine-related fake news by social media platform during 2024.}
    \label{fig:preliminary-social-platforms}
\end{figure}

The fact-checking articles also frequently identified the social media platforms where the false information was circulating. Figure~\ref{fig:preliminary-social-platforms} shows the number of verified claims per platform. Some articles indicated more than one platform circulating the fact-checked claim. In these cases, we considered all cited platforms. Because of that, the sum of the mentioned platforms (169) surpasses the number of collected articles (164). Instagram had the highest number of verified claims, followed by Facebook, X (formerly Twitter), Telegram, and WhatsApp. Among these top five platforms, Telegram was the only one with terms of service permissive to data collection for research purposes, provided that local regulations and data protection principles are respected. Attempts to collect data from other platforms were unfeasible due to restrictive data-sharing agreements or, in the case of X, prohibitively high costs for API access that exceeded our budget. Consequently, our data collection focused exclusively on the Telegram platform, targeting the most discussed and disseminated vaccine-related false information identified in our preliminary investigation.

The next section details our methodology for collecting and curating data from Telegram.

\section{Data Collection and Curation}
\label{sec:collection}

\subsection{Data Source and Selection}
Our preliminary investigation highlighted the importance of collecting data beyond healthcare-specific themes. Therefore, our scope included content related to politics, public figures, conspiracy theories, and scientific studies that contribute to the broader context of vaccine-related informational disorder. In the following, we detail the strategies we used to locate potential Telegram channels under the scope of vaccine information disorder.

\subsection*{Source Selection Strategy}
To identify Telegram channels spreading vaccine misinformation, we adopted keyword searches, a seed list of channels, and snowball sampling similar to that of Silva~\cite{Silva2024AntivaxTelegram}. We also limited the collection to public channels and groups that did not have restrictive entry policies and that had at least 1,000 members.

We began by inspecting a seed list of public Brazilian Telegram channels known for circulating anti-vaccine content. This list was compiled from the work of Monari~\cite{Monari2024FakeNews}, our own preliminary investigation with the fact-checking agencies, and additional sources from Silva~\cite{Silva2024AntivaxTelegram}. It is important to note that Telegram groups are dynamic; they can be deleted, renamed, or blocked by Brazilian authorities. This was noticeable when attempting to locate channels reported in prior works that were no longer active on the Telegram platform.

Following the approach of Silva~\cite{Silva2024AntivaxTelegram}, we expanded our list by including channels recommended by Telegram as ``similar'' to our seed channels. While official documentation on Telegram's similarity algorithm is unavailable, an inspection of these recommendations revealed that they often discussed similar topics and shared a similar audience with the seed channel.

\subsection*{Keyword-Based Search}
In addition to the channel-based selection, we performed a keyword-based search to identify further relevant groups. The search terms, designed to capture common anti-vaccine and conspiratorial language in Brazilian Portuguese, included: ``Vacina'', ``V@c1n@'', ``Picadinha'', ``Picada Veneno'', ``Injetar Veneno'', ``Veneno injetável'', ``mRNA'', ``COVID'', ``Negacionista'', ``Ignorante'', ``Passaporte vacinal'', ``Comprovante vacinal'', ``Experimento'', ``Vacina experimental'', ``mRNA'', ``Liberdade'', ``Eventos adversos'', ``Efeitos adversos'', ``Proteína Spike'', ``Detox vacinal'', ``Tratamento precoce'', ``Ivermectina'', ``\$iêb\&oa'', ``{\$iên\$}ia'', and ``NOM (Nova Ordem Mundial)''.

\subsection*{Collection Timeframe}
To understand the dynamics of vaccine-related information sharing, we established a collection period from \textbf{January 2020 to June 2025}. This timeframe was chosen to cover the initial circulation of anti-vaccine narratives in Brazil at the beginning of the COVID-19 pandemic through to the present.

Next, we detail the tools used for data collection.

\subsection{Data Collection Tool}

We developed a custom data collection tool in Python, built upon the open-source Telethon library~\cite{LonamiWebs2025Telethon}. Telethon facilitates interaction with the Telegram API via the MTProto protocol, allowing us to programmatically collect public data from the groups and channels identified in our search phase. The library provides a metadata-rich output and enables the collection of media content, including images, audio, videos, and polls.

\begin{figure}[ht!]
    \centering
    \includegraphics[width=0.8\textwidth]{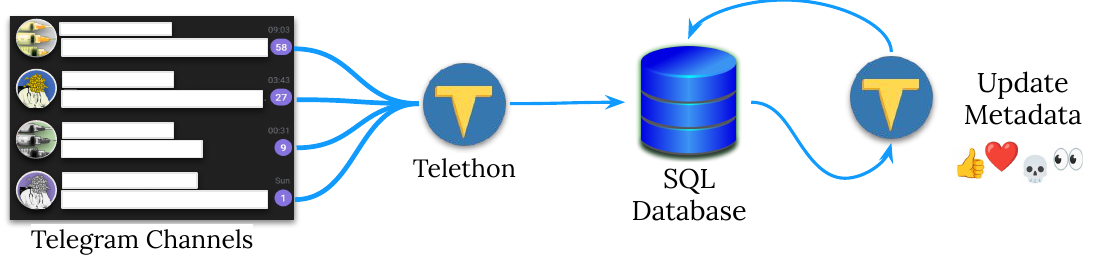}
    \caption{Data collection pipeline. Telethon collects information for each monitored channel, saves it on a SQL database, and updates its engagement metrics up to seven days after its initial collection.}
    \label{fig:data-collection-pipeline}
\end{figure}

Figure~\ref{fig:data-collection-pipeline} presents an overview of the tool's architecture. For each selected channel, the tool collected messages, media, and associated metadata within the predefined timeframe (January 2020 to June 2025). The collected content was then stored in a Structured Query Language (SQL) database.

Since engagement metrics such as views, reactions, and forwards evolve over time, we implemented a routine to update the metadata for each collected post for up to seven days after its initial collection.

\subsection{Data Acquisition}
Using the developed tool under the methodology described in the earlier sections, we collected data from \textbf{119 channels} totaling \textbf{3,998,633 Telegram posts}, of which \textbf{1,440,498 are associated with media} content (such as images, videos, audios, documents, and polls), totalizing \textbf{5.5 TB} of data. We restricted our data acquisition to media or documents with less than 50Mb. We set this constraint to avoid retrieving too large files that could overload our data storage. Table~\ref{tab:media_summary} describes the amount of collected data per type. 

\begin{table}[ht]
\centering
\caption{Distribution of collected media}
\begin{tabular}{l r}
\hline
\textbf{Media Type} & \textbf{Collected items} \\
\hline
Image & 765,750 \\
Video & 645,949 \\
Audio & 17,376 \\
Poll & 1,931 \\
Others & 9,490 \\
\hline
\textbf{Total} & \textbf{1,440,498} \\
\hline
\end{tabular}
\label{tab:media_summary}
\end{table}

\subsection{Data Curation and Processing}
After collecting the dataset, we processed the posts to identify their language and to indicate whether they are directly related to vaccine topics or not.
Our main goal with the curation is to facilitate the future analysis of data related to vaccines in the Portuguese language.

\subsubsection*{Text language identification}
\begin{figure}[ht!]
    \centering
    \includegraphics[width=0.8\textwidth]{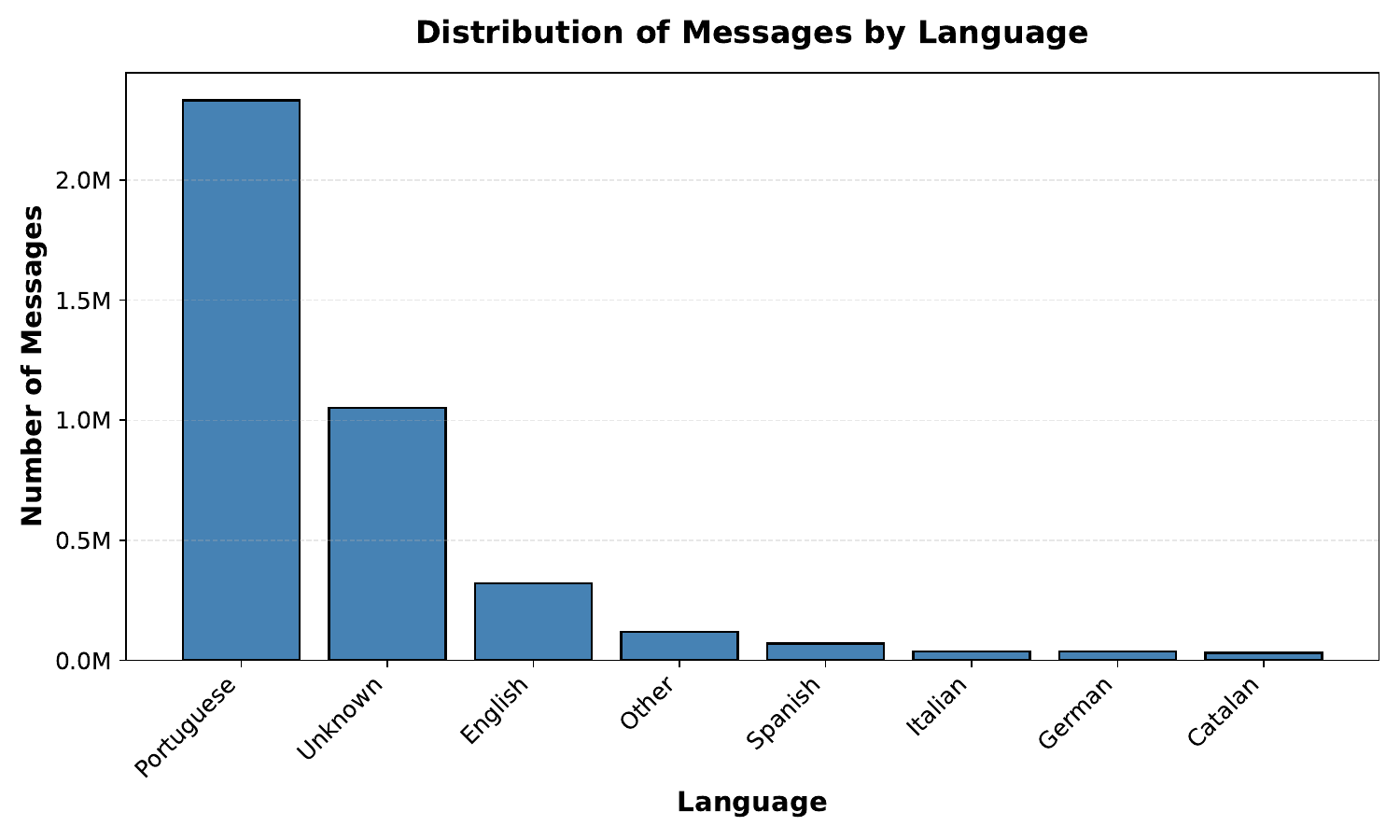}
    \caption{Language distribution of the collected posts}
    \label{fig:lang-dist}
\end{figure}

To identify the language of the posts, we use \textit{langdetect}~\cite{langdetect}. This library can identify multiple text languages using language profiles trained on the distribution of the n-gram character unique to each language. During inference, we set a minimum threshold of 0.5 for confidence in the detection. One million fifty-one thousand three hundred thirty posts (26\% of the data) were classified as Unknown due to low confidence or a lack of context to classify the posts. For example, posts that contain only URLs. Figure~\ref{fig:lang-dist} presents the distribution of the identified languages. According to \textit{langdetect}, most of the collected data is in Portuguese, followed by English and Spanish.

\subsubsection*{Vaccine-related posts identification}
To indicate whether a post discusses vaccines or not, we use the Large Language Model Sabiá-3~\cite{abonizio2024sabia3}, with the prompt from Table~\ref{tab:prompt-sabia}.

\begin{table}[htbp]
\centering
\caption{Prompt in Portuguese used in the LLM experiment.}
\vspace{0.3em}

\begin{tcolorbox}[
    colback=gray!10,
    colframe=gray!50,
    sharp corners,
    boxrule=0.4pt,
    width=\textwidth,
    listing only,
    breakable,
    listing options={
        basicstyle=\ttfamily\small,
        breaklines=true,
        columns=fullflexible,
        literate=
         {á}{{\'a}}1 {Á}{{\'A}}1
         {à}{{\`a}}1 {À}{{\`A}}1
         {ã}{{\~a}}1 {Ã}{{\~A}}1
         {â}{{\^a}}1 {Â}{{\^A}}1
         {é}{{\'e}}1 {É}{{\'E}}1
         {ê}{{\^e}}1 {Ê}{{\^E}}1
         {í}{{\'\i}}1 {Í}{{\'I}}1
         {ó}{{\'o}}1 {Ó}{{\'O}}1
         {ô}{{\^o}}1 {Ô}{{\^O}}1
         {õ}{{\~o}}1 {Õ}{{\~O}}1
         {ú}{{\'u}}1 {Ú}{{\'U}}1
         {ç}{{\c{c}}}1 {Ç}{{\c{C}}}1
    }
]
Você é um especialista em análise de texto e saúde pública. Sua tarefa é classificar se um texto está relacionado ao tópico de vacinas ou vacinação.

Critérios para classificar como RELACIONADO A VACINAS:
- Menciona vacinas, vacinação, imunização
- Discute efeitos, eficácia ou segurança de vacinas
- Fala sobre políticas de vacinação
- Contém teorias conspiratórias sobre vacinas
- Discute hesitação vacinal ou movimento antivacina

Responda APENAS "SIM" se o texto é relacionado a vacinas 
ou "NÃO" se não é relacionado, sem comentários.
\end{tcolorbox}
\label{tab:prompt-sabia}
\end{table}

We validate this prompt on a sample of 600 posts annotated by three experts on vaccine misinformation spreading. From these amounts, 282 were labeled as vaccine-related and 318 as not vaccine-related. They agreed on an average of 0.866 using the Cohen's Kappa metric.
Sabiá-3 using the presented prompt resulted in $90\%$ of F1-score in these samples. 

We were satisfied with the performance of the method and applied it to the remaining part of the dataset.
For posts with few words (e.g., less than 5 words), Sabiá-3 was not able to provide an answer, leaving these posts without a label. Table~\ref{tab:vaccine-annotation} indicates the amount of annotated data and ignored data by Sabiá-3.
We annotated the dataset using its non-anonymized text version to provide more context to Sabiá-3.

\begin{table}[ht]
\centering
\caption{Vaccine-related data annotation using the Large Language Model Sabiá-3}
\begin{tabular}{l r}
\hline
\textbf{Data Annotation} & \textbf{Number of Posts} \\
\hline
Vaccine-related & 407,723 \\
Not related to Vaccine & 2,222,909 \\
Not Annotated (ignored by Sabiá-3) & 1,368,001 \\
\hline
\textbf{Total} & \textbf{3,998,633} \\
\hline
\end{tabular}
\label{tab:vaccine-annotation}
\end{table}

\section{Data Description and Structure}
\label{sec:description}
The collected text posts and their metadata are available in a JSON Lines (.jsonl) format, with each line representing a valid JSON object for one message. Due to ethical concerns and the size of the collected media, access to the associated media (images, videos, and audio files) requires signing an agreement that outlines responsibilities and research ethics. This agreement ensures that the data will not be shared or used for non-research purposes, given the sensitivity of the information.
In the following, we describe the data scheme provided by the text posts datasets and some statistics regarding the dataset

\subsection{Data Schema}
Each message adheres to the schema described in Table \ref{tab:schema}. 
Some fields in the schema contain specific nuances, which are described in greater detail below.
\begin{itemize}
     \item \texttt{n\_forwards}: Indicates how many times the message has been forwarded to other channels. This metric reflects the message's propagation potential as a source of shared content.

    \item \texttt{forward\_from}: Identifies the original source (channel or user) from which the message was forwarded. This field shows that the message did not originate in the current channel but was shared from another feed.

    \item \texttt{forward\_from\_n\_forward}: Records the total number of times the original message (in the source channel) has been forwarded across Telegram.

    \item \texttt{views}: Represents the number of times a message has been viewed. According to Telegram's FAQ~\cite{telegram_faq_eye_icon}, this value is an approximation: if a user views the same message multiple times within four days, it counts as a single view; after that period, new views are counted again.

    \item \texttt{forward\_from\_views}: Indicates the cumulative number of views associated with the forwarded message, including those from all channels that have shared it. As noted in the Telegram FAQ~\cite{telegram_faq_eye_icon}, ``views from forwarded copies of a message are also included in the total count.'' This metric helps assess the overall reach of the message beyond its original source.

    \item \texttt{reply\_to}: If the message is a reply, this field stores the identifier (\texttt{message\_id}) of the original message being replied to within the same channel.
\end{itemize}

\begin{table}[htbp]
    \centering
    \caption{Schema of the dataset: structure and meaning of each message field.}
    \label{tab:schema}
    \renewcommand{\arraystretch}{1.2}
    \setlength{\tabcolsep}{5pt}

    \begin{tabular}{@{}p{5cm}p{2.2cm}p{9cm}@{}}
        \toprule
        \textbf{Field Name} & \textbf{Type} & \textbf{Description} \\ 
        \midrule

        \texttt{message\_id} & String & Unique identifier for the message within its chat. \\
        \texttt{channel\_id} & String & Unique identifier for the source channel or group. \\
        \texttt{user\_id} & String & SHA-256 hash of the author’s user ID (anonymized). \\

        \texttt{date} & Datetime & UTC timestamp indicating when the message was sent. \\
        \texttt{collected\_date} & Datetime & UTC timestamp indicating when the message was collected. \\
        \texttt{edit\_date} & Datetime & UTC timestamp of the message’s last edit. \\

        \texttt{text\_content} & String & Raw text of the message. Empty for non-text messages. \\
        \texttt{reply\_to} & String & Identifier of the parent message if this message is a reply. \\
        \texttt{n\_forwards} & Integer & Number of times the message has been forwarded. \\
        \texttt{views} & Integer & Number of times the message has been viewed. \\
        \texttt{reactions} & Integer & Total number of reactions received, regardless of type. \\

        \texttt{forward\_from} & String & Source channel if the message is a forward. \\
        \texttt{forward\_from\_views} & Integer & View count of the original forwarded message. \\
        \texttt{forward\_from\_reaction} & Integer & Reaction count of the original forwarded message. \\
        \texttt{forward\_from\_n\_forwards} & Integer & Forward count of the original forwarded message. \\

        \texttt{media\_type} & String & Type of media attached (e.g., \textit{document}, \textit{image}, \textit{video}). \\
        \texttt{media\_url} & String & URL associated with the media, if present. \\
        \texttt{media\_title} & String & Title of the media, if provided. \\
        \texttt{media\_description} & String & Description of the media (as provided by Telegram). \\

        \texttt{is\_vaccine\_related} & Boolean & Indicates whether the message concerns vaccine-related topics. \\
        \texttt{language} & String & Predominant language of the message (e.g., Portuguese, English). \\

        \bottomrule
    \end{tabular}
\end{table}

\subsection{Descriptive Statistics}
\begin{figure}[ht!]
    \centering
    \includegraphics[width=0.8\textwidth]{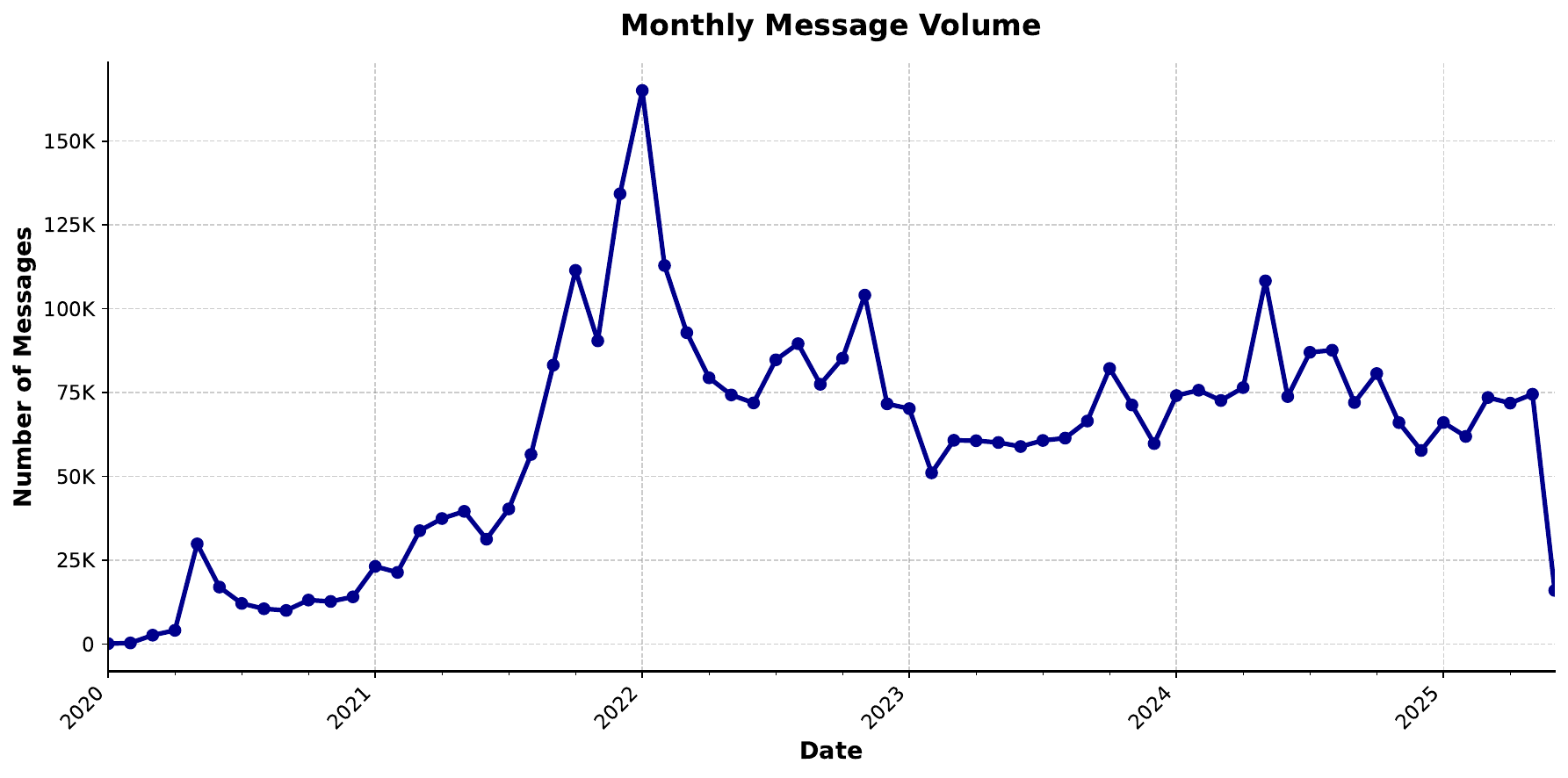} % Using a placeholder image
    \caption{Number of messages collected per month during the data collection period. The last point represents June of 2025, after which we stopped collecting the data.}
    \label{fig:temporal}
\end{figure}

Table \ref{tab:stats} provides an overview of the dataset's key statistics. A visual representation of the message volume over time is shown in Figure \ref{fig:temporal}.

\begin{table}[ht!]
    \centering
    \caption{Key statistics of the collected dataset.}
    \label{tab:stats}
    \begin{tabular}{@{}lr@{}}
        \toprule
        \textbf{Statistic} & \textbf{Value} \\
        \midrule
        Total Messages & 3,998,633 \\
        Total Channels/Groups & 119 \\
        Unique Anonymized Users & 71,672 \\
        Time Period Covered & Jan 2020 - Jun 2025 \\
        Messages with Text & 3,345,088 (83.6\%) \\
        Posts Replies & 1,048,850 (26.2\%) \\
        Forwarded Posts & 588,913 (14.7\%) \\
        Main Languages Detected & Portuguese (58.3\%), English (8.0\%), Spanish (1.7\%) \\
        Vaccine-Related & 407,723 (10.2\%)\\
        \bottomrule
    \end{tabular}
\end{table}

\section{Potential Uses and Applications}
\label{sec:uses}
This dataset was designed to identify and develop techniques to mitigate vaccine-hesitancy misinformation and disinformation circulating on social media. This can support a wide range of research endeavors across various fields:
\begin{itemize}
    \item \textbf{Natural Language Processing (NLP):} The large volume of conversational and social media text is suitable for training language models, sentiment analysis, topic modeling, and misinformation detection systems.
    \item \textbf{Social Science:} Researchers can study online community dynamics, the formation of echo chambers, and the spread of anti-vaccine narratives.
    \item \textbf{Network Analysis:} The forward and reply structure enables the construction of user interaction networks to identify influential community structures that could be useful to detect disinformation campaigns.
    \item \textbf{Synthetic Realities}: The media associated with each post can be used to identify generative artificial intelligence videos and images that are being used to support anti-vaccine ideas. 
\end{itemize}

\section{Data Availability}
\label{sec:availability}
The dataset is publicly available for non-commercial research purposes under the Creative Commons BY-NC 4.0 license. It is permanently hosted on Unicamp Institutional Research Data Repository (REDU) to ensure long-term availability it can be accessed via the following DOI:
\begin{center}
    \href{10.25824/redu/5JIVDT}{https://doi.org/10.25824/redu/5JIVDT}
\end{center}

\subsection{Ethical Considerations and Anonymization}
The data collection adheres to ethical guidelines, and it was approved by the Ethical Committee of the Universidade Estadual de Campinas (UNICAMP) (CEP/UNICAMP - 5404 and CAAE: 87658425.0.0000.5404). Only data from public channels and groups, accessible to any Telegram user, were collected. To protect user privacy, we performed the following anonymization steps:
\begin{itemize}
    \item \textbf{Channel Anonymization:} Channels and user IDs were replaced with a persistent, non-reversible SHA-256 hash, salted with a private key following standard guidelines~\cite{ENISA2019Pseudonymisation}. This prevents deanonymization while still allowing researchers to track these entities within the dataset.
    \item \textbf{PII Removal:} We use the open-source Presidio~\cite{microsoft_presidio} from Microsoft to detect and redact potential Personally Identifiable Information (PII) such as phone numbers, emails, and addresses from the message text.
\end{itemize}
Before anonymization, the dataset contained 97,506 unique user IDs and included timestamped entries for users joining and leaving channels. This timestamp information was identified as a privacy risk because it could be used to re-identify individuals (e.g., by correlating it with external knowledge of a user's join time). To mitigate this risk, we removed all join and leave entries. As a consequence, users whose only recorded activities were joining or leaving a channel were also excluded from the dataset. This step resulted in a final, anonymized dataset comprising 71,672 unique user IDs.

The collection and anonymization process was designed to comply with Telegram's Terms of Service and data protection regulations, such as the Brazilian General Personal Data Protection Law.

\section{Limitations}
\label{sec:limitation}
% It's important to be transparent.
% Mention potential biases:
% 1. Collection bias (your channel selection method may have missed things).
% 2. Platform bias (Telegram users are not representative of the general population).
% 3. Data quality (spam, bots, imperfect PII removal).
% \subsection{Challenges and Contingencies in Data Collection}
During the data collection, we faced different challenges and unpredictable behavior in collecting and safeguarding the integrity of the data. We enumerated them in the following:

\begin{enumerate}
    \item \textbf{Telegram API rate limit}: To use the Telegram API, we registered an account created specifically for this research project within the API developer environment, ensuring compliance with its terms of service. This required adherence to API request limits, which the Telethon library automatically managed. To manage these rate limits, we implemented a backoff mechanism that paused the collector for a variable duration, as determined by the Telethon library, before resuming operations. This waiting period could range from seconds to several hours, depending on the volume of data collected.

    This rate-limiting, combined with the substantial time required to download large media files (images and videos), extended the total data collection period to five months. Future work would benefit from exploring a solution to accelerate this process by orchestrating multiple Telegram accounts, each associated with a unique phone number, to speed up that collection while avoiding rate limits.

    \item \textbf{Storage capacity}: To avoid overloading our storage capacity, we only collected media with at most 50Mb. Any file shared on the monitored channels with a size of 50 MB or higher was ignored during data collection.

    \item \textbf{Engagement Metrics}: While collecting the data, we noticed that the post reaction feature (e.g., likes) was implemented in Telegram only after December 30, 2021. Because of that, there is a gap in such metadata on content posted before that date.

    \item \textbf{Added to groups without consent}: As we were monitoring and collecting the data of all groups associated with our Telegram account, we noticed that our account was automatically added to channels without our consent, so we had to discard the collected data from these channels later. Most of these channels were related to Bitcoin information exchange or product selling.

    \item \textbf{Dynamic Channels}: During data collection, we noticed some of the monitored channels were deleted or changed their names, which forced us to track their channel IDs or the new channel they moved to.
    
    \item \textbf{Public channels changing to private}: A few channels that we collected changed their behavior during data collection, which initially were public groups but changed to private chats. This prevents us from accessing their data from that moment on.

    \item \textbf{Channels infringing Brazilian laws}: Due to fake information shared by the monitored channels, we noticed that multiple messages were deleted by the Brazilian Supreme Court. When this happens, the content of the message is changed to a statement that the message was removed by the Supreme Court due to legal infringement. We also noticed that some monitored channels were blocked or deleted for the same reason.
\end{enumerate}

\section{Conclusion}
\label{sec:conclusion}
% Summarize the paper, reiterate the value of your dataset, and encourage its use by the research community.
The dataset presented arises from the need to better understand the dynamics of information disorder circulating within Brazilian online anti-vaccine communities.
After analyzing fact-checking sources and identifying social media platforms that facilitate the spread of misinformation, we focused on collecting posts from Telegram channels known for disseminating misleading content related to healthcare and vaccines.

Our dataset encompasses messages from 119 public channels relevant to the Brazilian anti-vaccine ecosystem, forming a unique corpus of approximately four million posts. Notably, at least 10\% of these messages directly relate to vaccine discussions. This collection provides a valuable foundation for the scientific community to explore, in depth, how misinformation emerges, how it is shared within communities, and which narratives contribute to vaccine hesitancy.

Beyond its research value, we hope this work contributes to building more informed, empathetic, and resilient public health communication. By understanding misinformation patterns, governments, health authorities, and civil society can design strategies that not only counter false narratives but also respond with compassion, respect, and care toward those influenced by them. Ultimately, fostering dialogue and trust is as essential as combating misinformation, protecting both individuals and communities through shared understanding.

The Brazilian Social Media Anti-vaccine Information Disorder
Dataset is available at \url{https://doi.org/10.25824/redu/5JIVDT}, under Creative Commons BY-NC 4.0 license.

\section*{Acknowledgments}
We gratefully acknowledge \textbf{Maritaca.ai} for generously providing computational credits and assistance with data filtering used in this study.
We thank the São Paulo Research Foundation (\textbf{FAPESP}) under the Horus Project (\#2023/12865-8), and the National Council for Scientific and Technological Development (\textbf{CNPq}), and the Department of Science and Technology of the Secretariat of Science, Technology and Innovation and the Economic-Industrial Health Complex of the \textbf{Ministry of Health of Brazil} under the Aletheia Project (\#442229/2024-0), for their financial support of this research.

%======================================================================================
% BIBLIOGRAPHY
%======================================================================================
\printbibliography

\end{document}